%% file: paper.tex
\documentclass{sig-alternate-05-2015}

\usepackage{cite}
\usepackage{url}
\usepackage{graphicx}
\usepackage{booktabs}
\usepackage{tabularx}
\usepackage{multirow}
\usepackage{subfig}
\usepackage{amsmath,dsfont}
\usepackage{array}
\usepackage{fixltx2e} 

\usepackage{algorithm}
\usepackage{algorithmic}

\usepackage{pifont}

\usepackage{verbatim}

\usepackage{colortbl}
\definecolor{mygray}{gray}{.75}

\usepackage{color}
\usepackage{amssymb}
\usepackage[normalem]{ulem}

\usepackage{adjustbox}

\begin{document}


\doi{}

\isbn{}



%

\title{Detecting Violence in Video using Subclasses}
%
%
%
%
%

\numberofauthors{1} 
%
\author{
%
%
\alignauthor
Xirong Li, Yujia Huo, Jieping Xu, Qin Jin\\
\affaddr{Multimedia Computing Lab, Renmin University of China}\\
}

\maketitle
\begin{abstract}

This paper attacks the challenging problem of violence detection in videos.
Different from existing works focusing on combining multi-modal features,
we go one step further by adding and exploiting subclasses visually related to violence.
We enrich the MediaEval 2015 violence dataset by \emph{manually} labeling violence videos with respect to the subclasses.
Such fine-grained annotations not only help understand what have impeded previous efforts on learning to fuse the multi-modal features,
but also enhance the generalization ability of the learned fusion to novel test data.
The new subclass based solution,
with AP of 0.303 and P100 of 0.55 on the MediaEval 2015 test set,
outperforms several state-of-the-art alternatives.
Notice that our solution does not require fine-grained annotations on the test set,
so it can be directly applied on novel and fully unlabeled videos.
Interestingly, our study shows that motion related features, though being essential part in previous systems, are dispensable. 
\end{abstract}

%
%


%
%

%
%



\section{Introduction} \label{sec:intro}

\input{intro}

\section{Subclass Annotation} \label{sec:anno}

\input{subclass}

\section{Using Subclasses for Violence Detection} \label{sec:main}

\input{setup}

\input{eval}

\section{Summary and Conclusions} \label{sec:conc}

We contribute to violence video detection as follows.
First, we enrich the MediaEval 2015 violence dataset by providing video-level annotations with respect to ten subclasses of violence.
The clear divergence between subclass distributions of the MediaEval dev and test sets explains why existing learn-to-fuse attempts did not work.
Second, by a systematic comparison between violence detection with and without subclasses,
we re-justify the effectiveness of learn-to-fuse for violence video detection.
The new solution beats several state-of-the-art systems on the MediaEval 2015 test set.
Moreover, decomposing violence into subclasses results in novel findings concerning complementariness of distinct modalities.
The conclusion of \cite{mtap2016-vsd-vlam} that visual and motion features are better than audio features holds when modeling violence in a holistic manner.
For fusion of multiple subclasses and multi-modal features, motion related features are in fact ignorable.


%
\bibliographystyle{abbrv}
\bibliography{paper}  
%
%

\end{document}

%% file: intro.tex

\begin{figure} [tb!]
\centering
\includegraphics[width=\columnwidth]{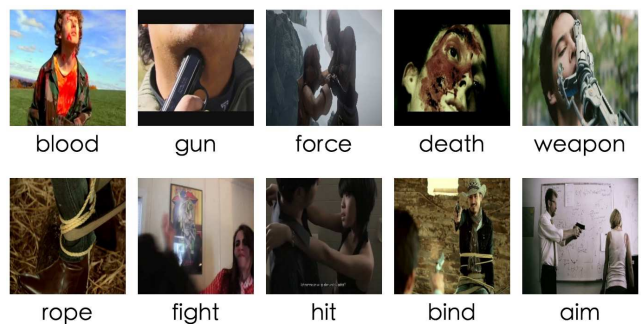}
\caption{Visual examples of ten violence subclasses.
We enrich the MediaEval 2015 violence data \cite{vsd2015-overview} by providing video-level annotations with respect to these subclasses.} 
\label{fig:visual-examples}
\end{figure}

Detecting violence in videos is important for protecting children against offensive content. 
Viewing violence as a specific concept, one might treat violence detection as a showcase of video concept detection, 
an extensively studied topic in the literature \cite{lscom,ye2015}.
The MediaEval affective task defines violence as \textit{one would not let an 8 years old child see because of their physical violence} \cite{vsd2015-overview}.
Scenes of violence lack consistent visual patterns, see Fig. \ref{fig:visual-examples}.
Moreover, the perception of violence varies over subjects.
Another evidence showing the different landscape of violence detection 
is that state-of-the-art video concept detection uses deep image features alone \cite{med2015:uva}.
By contrast, a violence detection system often involves a number of multi-modal features 
related to image, audio and motion \cite{mtap2016-vsd-vlam,mediaeval2015-fudan,mediaeval2015-tju}.
We also use multi-modal features,
but go one step further by adding and exploiting subclasses visually related to violence,
for better performance and deeper insight.


\textbf{Progress on video violence detection}.
Thanks to the MediaEval benchmark effort \cite{mediaeval2013,mtap2015-vsd,vsd2015-overview}, 
data for violence detection research has been substantially expanded in terms of types of violence and the amount of videos,
compared to earlier work on a specific type of violence, e.g., fight in hockey games \cite{caip2011-violence-detection}.
Consequently, different from previous works that mainly utilize motion related features such as motion SIFT and histogram of optical flow \cite{caip2011-violence-detection,mo_sparse},
current solutions embrace multi-modal features to describe video content in terms of visual appearance, audio channel and motion over the temporal dimension. 
A common package, as reported in \cite{mtap2016-vsd-vlam,mediaeval2015-fudan,mediaeval2015-tju,mediaeval2015-ruc}, 
consists of image features extracted by pre-trained deep convolutional neural networks \cite{vggnet,googlenet},
MFCC based audio features \cite{specom2005-mfcc}, and local descriptors derived from improved dense trajectories \cite{iccv2013-idt}.

As for combining the many features, although late fusion is consistently found to improve over models derived from single features \cite{mediaeval2015-nii,mediaeval2015-tju,mediaeval2015-ruc}, 
the success so far has been limited to average weighting, i.e., all the underlying models are treated equally.
As the effectiveness of the features varies \cite{mtap2016-vsd-vlam}, average fusion is unlikely to be the optimal solution.
Nonetheless, efforts on learn-to-fuse are found to be inferior to the average-fusion baseline \cite{mediaeval2015-ruc}.
While current works simply attribute this to divergence between training and test data \cite{mediaeval2015-ruc,mediaeval2015-nii},
we believe a more in-depth analysis is required to understand what impedes the use of learn-to-fuse for video violence detection.

\medskip

\textbf{Contributions of this work}.
In order to better understand the difficulty in generalizing models learned on a given dataset to a novel test set,
we first enrich the MediaEval 2015 violence set, an up-to-date benchmark for video violence detection, 
by providing video-level manual annotations with respect to ten subclasses of violence.
We are inspired by the work from Tan and Ngo \cite{mediaeval2013-vireo}, 
which makes an interesting attempt to express violence by a number of middle-level concepts.
Nonetheless, as they target at learning these concepts from YouTube videos to derive a semantic feature,
the two works are driven by completely different motivations.
As we will show in Section \ref{sec:anno}, the newly added subclass annotations help interpret cross-dataset divergence 
and consequently the difficulty of generalizing the learned fusion to previously unseen test data.
Moreover, we show in Section \ref{sec:main} how this difficulty can be resolved to some extent using subclass based violence detection,
and learned fusion of multiple features and subclasses outperforms several state-of-the-art alternatives.

%% file: subclass.tex
\textbf{Video sets}.
We use the official release of the MediaEval 2015 affective task \cite{vsd2015-overview}.
Containing 10,900 short video clips extracted from 199 professional and amateur movies, 
this benchmark dataset is much larger and more diverse than those narrow-domain videos used in previous works  \cite{caip2011-violence-detection,mo_sparse}.
The MediaEval dataset consists of two disjoint subsets: a dev set of 6,144 clips and a test set of 4,756 clips.
Notice that the videos were not originally selected for violence, so they have low occurrence of violence, with 4.4\% in the dev set and 4.8\% in the test set.

\textbf{Manual labeling procedure}.
We aim to provide fine-grained annotations of violence videos by manually assigning visual concepts that are highly relevant to violence.
Such a concept is supposed to have relatively consistent visual appearance and thus more easy to be detected.
We take the 52 middle-level concepts from \cite{mediaeval2013-vireo} as our initial concept vocabulary.
Given the subjective definition of video violence, the vocabulary may expand with novel concepts as the manual labeling process goes on.
We watched all violence videos in the dev set, eventually yielding a vocabulary of 95 concepts.
For a specific concept, there shall be a reasonable amount of positive instances, otherwise modeling this concept is very likely to be futile.
We empirically set the threshold to be 20, i.e., a concept needs to have over 20 occurrences in the dev set.
This results in ten subclasses of violence, as exemplified in Fig. \ref{fig:visual-examples}.
Four of them, namely `bind', `aim', `weapon' and `death', are not covered by the initial 52 concepts.
A video can be labeled with multiple subclasses.
Some subclasses tend to co-occur with some others in the same video clips, e.g., `rope' to `bind' and `blood' to `death'. 
However, `death' and `fight' seldom co-occur, as a fight hardly ends with death in this dataset.

\begin{figure}[tb!]
\centering
\includegraphics[width=\columnwidth]{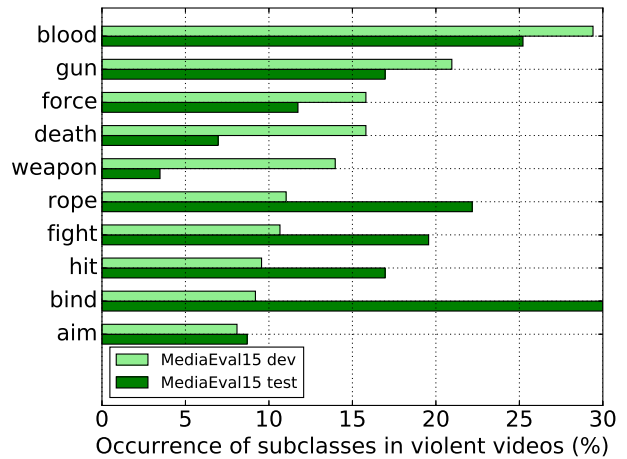}
\caption{Occurrence of each subclass in violence videos, clearly varying across datasets.
Such divergence adds difficulty of generalizing models trained on the dev set to the previously unseen test set.}
\label{fig:subclass-dist}
\end{figure}

In order to reveal potential divergence between the dev set and the test set, 
we repeat the above procedure to annotate all violence videos in the test set.
Fig. \ref{fig:subclass-dist} shows the occurrence rate of each subclass, which clearly varies across the two datasets.
Such divergence suggests the challenge of applying a fusion model trained on the dev set to the (unseen) test set.

Note that the fine-grained annotations of the test set are meant for cross-dataset analytics only.
They are not used for violence detection.
So, similar to existing works \cite{mtap2015-vsd,mediaeval2015-fudan,mediaeval2015-tju},
our subclass based violence detection solution is directly applicable to novel video data that are completely unlabeled.

%% file: setup.tex
In order to justify the viability of subclass based violence detection,
we build two systems, with and without subclasses, respectively.
We setup the systems in Section \ref{ssec:setup}, followed by experiments in Section \ref{ssec:exp}.

\subsection{Setup} \label{ssec:setup}

Following the current practice \cite{mediaeval2015-fudan,mediaeval2015-tju,mediaeval2015-nii},
we employ multi-modal video representations including image features extracted by deep convolutional neural networks\cite{googlenet,vggnet},
audio features \cite{specom2005-mfcc}, and motion features \cite{iccv2013-idt}.
They are described as follows, with a quick summary given in Table \ref{tab:feat}.

\textbf{Image CNN features}.
Given a video, we uniformly extract its frames with an interval of 0.5 second,
resulting in 131,441 and  101,587 frames in the dev and test sets, respectively.
For each frame we extract deep features using three pretrained CNN models,
i.e., VGGNet \cite{vggnet}, GoogleNet \cite{googlenet}, 
and GoogletNet4k \cite{googlenet4k}, a very recent variant of GoogleNet trained on a bottom-up reorganization of the ImageNet hierarchy.
Consequently, 
we obtain three frame-level features by taking the last fully connected layer of VGGNet and the pool5 layer of the two GoogleNets.

A video-level representation can significantly reduce the amount of training / test examples and is commonly used in video event detection \cite{ye2015,xu2015discriminative}. We obtain three video-level features by average pooling on the corresponding frame-level features.

\textbf{Audio features}.
We use the Mel-frequency Cepstral Coefficients (MFCCs) \cite{specom2005-mfcc}, 
computed over a sliding short-time window of 25ms with a 10ms shift. 
An audio segment is then represented by a set of MFCC features. 
In order to transform these features into a fixed-dimension vector, 
we adopt two encoding schemes, i.e., bag-of-audio-words by hard assignment and Fisher Vector encoding \cite{fisher}.

\textbf{Motion features}.
The Improved Dense Trajectory (IDT) \cite{iccv2013-idt} is used to capture actions in a video.
In particular, we use three trajectory based descriptors, 
including Motion Boundary Histogram (MBH), Histogram of Oriented Gradient (HOG), and Histogram of Optical Flow (HOF).
In a similar manner to the audio features, 
we use bag-of-words and Fisher Vector encodings to convert each of the descriptors into a fixed-dimension feature vector,
resulting in six motion related features.

\begin{table}[tb!]
  \renewcommand{\arraystretch}{1.2}
  \centering
  \caption{Fourteen features used in our experiments, 
  describing video content in varied aspects.}
  \label{tab:feat}%
  \scalebox{0.9}{
     \begin{tabular}{@{}|l|l|l|r|@{}}

    \toprule
    
    \textbf{Modality} & \textbf{Feature} & \textbf{Notation} & \textbf{Dim.}  \\ 
    \midrule
 \multirow{6}*{\emph{Image}} &   frame-level vggnet     &  vnet\textsubscript{f}   & 4,096                  \\
    \cmidrule(lr){2-4}
                        & video-level vggnet        &  vnet\textsubscript{v}    & 4,096                  \\
    \cmidrule(lr){2-4}
                        & frame-level googlenet     &  gnet\textsubscript{f}    & 1,024                 \\
    \cmidrule(lr){2-4}
                        & video-level googlenet     &  gnet\textsubscript{v}    & 1,024                \\
    \cmidrule(lr){2-4}
                        & frame-level googlenet4k   &  g4k\textsubscript{f}     & 1,024               \\
    \cmidrule(lr){2-4}
                        & video-level googlenet4k   &  g4k\textsubscript{v}     & 1,024               \\
    \midrule
  \multirow{2}*{\emph{Audio} } &  mfcc + bow         & mfcc\textsubscript{b}   & 4,096             \\
                       \cmidrule(lr){2-4}
                        &  mfcc + fisher vector   & mfcc\textsubscript{fv}      & 19,968       \\
    \midrule
  \multirow{6}*{\emph{Motion} } &  mbh + bow         & mbh\textsubscript{b}    & 4,000   \\
                        \cmidrule(lr){2-4}
                        &  mbh + fisher vector   & mbh\textsubscript{fv}   & 98,304  \\
                        \cmidrule(lr){2-4}
                        &  hog + bow         & hog\textsubscript{b}    & 4,000   \\
                        \cmidrule(lr){2-4}
                        &  hog + fisher vector   & hog\textsubscript{fv}   & 49,152  \\
                        \cmidrule(lr){2-4}
                        &  hof + bow         & hof\textsubscript{b}    & 4,000   \\
                        \cmidrule(lr){2-4}
                        &  hof + fisher vector   & hof\textsubscript{fv}   & 55,296  \\
    \bottomrule

    \end{tabular}%
  }
  \end{table}%

\medskip

\textbf{Classification Models}. 
For each of the ten subclasses as well as the holistic violence class, we train binary linear SVMs classifiers.
To effectively learn from the large amounts of imbalanced examples,
we leverage the Negative Bootstrap algorithm \cite{tmm2013-xli},
which iteratively and adaptively finds negative training examples that better improve classification than negatives selected at random.
For each frame-level classifier, we obtain its prediction on a given video by first scoring every frame, smoothing the scores along the temporal dimension, and then taking the maximal response.

The dev set is randomly divided into two disjoint subsets, 70\% for training the SVM models and 30\% as a val set to optimize hyper-parameters.

\textbf{Strategies for multi-feature multi-subclass fusion}. 
We investigate late fusion with two weighting strategies.
One is average weighting, i.e., assigning equal weights to the base classifiers.
Despite its simplicity, the average strategy is found to be effective for multi-feature fusion for violence detection \cite{mtap2016-vsd-vlam}.
The second strategy is learning to fuse. 
In particular, we optimize the weights on the val set using coordinate ascent, 
which optimizes a chosen weight per iteration.
A recent study in the context of image classification \cite{tmm2015-tagfeature} shows that 
compared to more complicated fusion algorithms such as multiple kernel learning,
coordinate ascent is more effective by directly optimizing (non-differential) performance metrics.

Putting all the features and subclasses together, we need to learn weights for up to 140 classifiers.
By contrast, for the system without subclasses, we only need to optimize weights for 14 classifiers at maximum.
Hence, comparing the two systems shall allow us to conclude whether the introduction of subclasses makes sense.

\textbf{Evaluation criteria}. 
Following the protocol \cite{vsd2015-overview}, we report Average Precision (AP), Precision at the top 10 ranked videos (P10), and P100.

\begin{table}[tbh!]
\renewcommand{\arraystretch}{1.2}
\centering
\caption{Performance of the indiviudal features, sorted in descending order according to AP scores of the \emph{learn} run on the test set.
For each feature, top performers on a given dataset are shown in bold font.}
\label{tab:feat-perf}
\scalebox{0.8}{
\begin{tabular}{@{}|l|r|r|r|r|r|r|@{}}
\toprule    
 & \multicolumn{3}{c|}{\textbf{MediaEval15 val set}} & \multicolumn{3}{c|}{\textbf{MediaEval15 test set}}  \\ 
 \cmidrule(lr){2-4} \cmidrule(lr){5-7}
\textbf{Feature} & \emph{w/o subclass} & \emph{avg} & \emph{learn} & \emph{w/o subclass} & \emph{avg} & \emph{learn} \\
\midrule
g4k\textsubscript{f}&0.321&0.260&\textbf{0.365}&0.190&0.179&\textbf{0.216}\\
\midrule
g4k\textsubscript{v}&0.284&0.237&\textbf{0.292}&\textbf{0.213}&0.195&0.199\\
\midrule
vnet\textsubscript{v}&0.295&0.263&\textbf{0.316}&0.120&0.137&\textbf{0.156}\\
\midrule
gnet\textsubscript{f}&0.245&0.210&\textbf{0.316}&0.097&\textbf{0.164}&0.156\\
\midrule
hog\textsubscript{b}&0.164&0.132&\textbf{0.174}&0.160&\textbf{0.175}&0.153\\
\midrule
vnet\textsubscript{f}&\textbf{0.347}&0.226&0.339&0.118&\textbf{0.147}&0.145\\
\midrule
mfcc\textsubscript{fv}&0.313&0.270&\textbf{0.316}&0.100&\textbf{0.138}&0.136\\
\midrule
gnet\textsubscript{v}&0.296&0.258&\textbf{0.312}&\textbf{0.152}&0.142&0.130\\
\midrule
mbh\textsubscript{b}&0.119&0.109&\textbf{0.197}&0.122&\textbf{0.129}&0.092\\
\midrule
hof\textsubscript{b}&0.127&0.130&\textbf{0.170}&\textbf{0.113}&0.109&0.092\\
\midrule
mfcc\textsubscript{b}&\textbf{0.320}&0.234&0.295&\textbf{0.090}&0.075&0.077\\
\midrule
mbh\textsubscript{fv}&\textbf{0.199}&0.075&0.159&\textbf{0.099}&0.064&0.067\\
\midrule
hof\textsubscript{fv}&\textbf{0.134}&0.083&0.096&\textbf{0.102}&0.050&0.054\\
\midrule
hog\textsubscript{fv}&\textbf{0.225}&0.070&0.101&\textbf{0.078}&0.047&0.049\\
\bottomrule
\end{tabular}
}
\end{table}

\input{table-fusion}

%% file: table-fusion.tex
\begin{table*}[tbh!]
\renewcommand{\arraystretch}{1.2}
\centering
\caption{Performance of feature-subclass fusion, sorted in descending order according to AP scores of the \emph{learn} run on the test set.
For each feature-fusion setting, top performers on a given dataset are shown in bold font.}
\label{tab:feat-fusion}
\scalebox{0.72}{
\begin{tabular}{@{}|l|l|l|l|l| l| l|l|l|l| l|l|l|l |r|r|r|r|r|r|r|r|@{}}
\toprule    
 \multicolumn{14}{|c}{Varied settings of feature fusion} &  \multicolumn{4}{|c|}{MediaEval15 val set} & \multicolumn{4}{c|}{MediaEval15 test set} \\
 \midrule
\multicolumn{6}{|c}{Image CNN} & \multicolumn{2}{|c}{Audio MFCC } & \multicolumn{6}{|c}{Motion IDT} & \multicolumn{2}{|c|}{w/o subclass} & \multicolumn{2}{c|}{with subclass} & \multicolumn{2}{c|}{w/o subclass} & \multicolumn{2}{c|}{with subclass}\\
\midrule 
vnet\textsubscript{f}&vnet\textsubscript{v}&gnet\textsubscript{f}&gnet\textsubscript{v}&g4k\textsubscript{f}&g4k\textsubscript{v}&mfcc\textsubscript{b}&mfcc\textsubscript{fv}&mbh\textsubscript{b}&mbh\textsubscript{fv}&hog\textsubscript{b}&hog\textsubscript{fv}&hof\textsubscript{b}&hof\textsubscript{fv}  & \emph{avg} & \emph{learn} & \emph{avg} & \emph{learn} & \emph{avg} & \emph{learn} & \emph{avg} & \emph{learn} \\
\midrule
\ding{52}&\ding{52}&\ding{52}&\ding{52}&\ding{52}&\ding{52}&\ding{52}&\ding{52}&\ding{56}&\ding{56}&\ding{56}&\ding{56}&\ding{56}&\ding{56}&0.484&\textbf{0.511}&0.386&0.510&0.274&0.251&0.288&\textbf{0.303}\\
\midrule
\ding{56}&\ding{56}&\ding{56}&\ding{56}&\ding{52}&\ding{52}&\ding{52}&\ding{52}&\ding{52}&\ding{56}&\ding{52}&\ding{56}&\ding{52}&\ding{56}&0.376&0.492&0.293&\textbf{0.493}&0.269&0.275&0.277&\textbf{0.296}\\
\midrule
\ding{56}&\ding{56}&\ding{56}&\ding{56}&\ding{52}&\ding{52}&\ding{52}&\ding{52}&\ding{56}&\ding{56}&\ding{56}&\ding{56}&\ding{56}&\ding{56}&0.455&0.473&0.397&\textbf{0.491}&0.248&0.236&0.283&\textbf{0.290}\\
\midrule
\ding{52}&\ding{52}&\ding{52}&\ding{52}&\ding{52}&\ding{52}&\ding{52}&\ding{52}&\ding{52}&\ding{52}&\ding{52}&\ding{52}&\ding{52}&\ding{52}&0.442&\textbf{0.531}&0.279&0.522&0.248&0.270&0.226&\textbf{0.279}\\
\midrule
\ding{52}&\ding{52}&\ding{52}&\ding{52}&\ding{52}&\ding{52}&\ding{52}&\ding{56}&\ding{52}&\ding{56}&\ding{52}&\ding{56}&\ding{52}&\ding{56}&0.440&\textbf{0.527}&0.282&0.511&\textbf{0.300}&0.260&0.259&0.275\\
\midrule
\ding{52}&\ding{52}&\ding{52}&\ding{52}&\ding{52}&\ding{52}&\ding{56}&\ding{56}&\ding{52}&\ding{56}&\ding{52}&\ding{56}&\ding{52}&\ding{56}&0.394&\textbf{0.482}&0.254&0.435&\textbf{0.280}&0.225&0.249&0.263\\
\midrule
\ding{52}&\ding{52}&\ding{52}&\ding{52}&\ding{52}&\ding{52}&\ding{52}&\ding{56}&\ding{56}&\ding{56}&\ding{56}&\ding{56}&\ding{56}&\ding{56}&0.468&0.495&0.353&\textbf{0.512}&\textbf{0.266}&0.228&0.250&0.263\\
\midrule
\ding{52}&\ding{52}&\ding{52}&\ding{52}&\ding{52}&\ding{52}&\ding{56}&\ding{56}&\ding{56}&\ding{56}&\ding{56}&\ding{56}&\ding{56}&\ding{56}&0.427&\textbf{0.449}&0.291&0.419&0.244&0.233&0.239&\textbf{0.255}\\
\midrule
\ding{52}&\ding{52}&\ding{52}&\ding{52}&\ding{52}&\ding{52}&\ding{56}&\ding{56}&\ding{52}&\ding{52}&\ding{52}&\ding{52}&\ding{52}&\ding{52}&0.373&\textbf{0.479}&0.226&0.447&0.229&0.249&0.202&\textbf{0.253}\\
\midrule
\ding{52}&\ding{52}&\ding{52}&\ding{52}&\ding{52}&\ding{52}&\ding{56}&\ding{56}&\ding{56}&\ding{52}&\ding{56}&\ding{52}&\ding{56}&\ding{52}&0.434&\textbf{0.468}&0.250&0.444&0.238&\textbf{0.253}&0.171&0.231\\
\midrule
\ding{56}&\ding{56}&\ding{56}&\ding{56}&\ding{56}&\ding{56}&\ding{52}&\ding{52}&\ding{52}&\ding{52}&\ding{52}&\ding{52}&\ding{52}&\ding{52}&0.308&0.424&0.194&\textbf{0.446}&\textbf{0.164}&0.141&0.141&0.141\\
\midrule
\ding{56}&\ding{56}&\ding{56}&\ding{56}&\ding{56}&\ding{56}&\ding{52}&\ding{52}&\ding{56}&\ding{56}&\ding{56}&\ding{56}&\ding{56}&\ding{56}&0.347&0.348&0.273&\textbf{0.353}&0.107&0.106&0.111&\textbf{0.117}\\
\midrule
\ding{56}&\ding{56}&\ding{56}&\ding{56}&\ding{56}&\ding{56}&\ding{56}&\ding{56}&\ding{52}&\ding{52}&\ding{52}&\ding{52}&\ding{52}&\ding{52}&0.206&0.279&0.146&\textbf{0.296}&\textbf{0.143}&0.111&0.116&0.114\\
\bottomrule
\end{tabular}
}
\end{table*}

%% file: eval.tex
\subsection{Experiments} \label{ssec:exp}

\textbf{Experiment 1. The impact of subclasses in a single-feature setting}.
Table \ref{tab:feat-perf} shows the performance of the individual features,
where \emph{w/o class} indicates runs without considering subclasses, 
\emph{avg} means average fusion of the ten subclasses,
and \emph{learn} is fusion with weights tuned on the val set.
In the \emph{w/o class} setting, 
the best motion feature hog\textsubscript{b}, with AP of 0.160, outperforms the best standard CNN feature, i.e., gnet\textsubscript{v}.
This is consistent with a recent result reported in \cite{mtap2016-vsd-vlam}.
Nonetheless, g4k\textsubscript{v} with AP of 0.213 is the top performer.

Half of the features improve by combining subclass classifiers,
with the best performance obtained by gnet\textsubscript{f} + \emph{learn}.
Interestingly, the majority of the features for which fusion does not work are motion related.
By looking in combinations of one subclass and one feature for violence detection,
we find that the motion features are in general less effective for modeling subclasses when compared with image CNN features.

\textbf{Experiment 2. The impact of subclasses in a multi-feature setting}.
We now move to the setting where multiple features are combined together with subclasses.
Given that there are over 16k combinations, we take a more practical approach 
by starting with using all the features.
Specific features are progressively removed with preference given to slower and longer features.
E.g., fisher vectors are removed in advance to their bag-of-words counterparts.
Moreover, in order to study complementariness of distinct modalities, 
features of an unchosen modality are left out.
Eventually we obtain 13 feature combinations with their performance reported in Table \ref{tab:feat-fusion}.

Comparing with the single feature setting, multi-feature fusion improves the performance.
In the \emph{w/o subclass} setting, however, fusion weights learned on the val set often yields lower AP than average weighting.
As shown in  Table \ref{tab:feat-fusion}, for only 4 out of 13 times \emph{learn} is better than \emph{avg}.
By contrast, the use of subclasses makes \emph{learn} beats \emph{avg} 11 out of 13 times.
This result shows the importance of subclasses for learning to fuse.

Concerning the complementariness of distinct modalities, 
all the modalities are required for deriving a holistic violence detector, 
as in the \emph{w/o subclass} scenario.
However, once subclasses are taken into account, the joint use of image and audio modalities (with motion excluded) performs the best.

Finally, we make a system level comparison between the proposed subclass based system and three state-of-art systems as demonstrated in the MediaEval 2015 evaluation. 
The result shown in Table \ref{tab:sota} further confirms the effectiveness of subclass based violence detection.

\begin{table}[tb!]
  \renewcommand{\arraystretch}{1.2}
  \centering
  \caption{Comparing with front-runners in the MediaEval 2015 violence detection task.}
  \label{tab:sota}%
  \scalebox{0.88}{
     \begin{tabular}{@{}|l|r|r|r|@{}}

    \toprule

    \textbf{System} & \textbf{AP} & \textbf{P10} & \textbf{P100} \\
    \midrule
    Fudan-Huawei \cite{mediaeval2015-fudan}   &  0.296     & 1.0           &  0.46        \\
    \midrule
    MIC-TJU \cite{mediaeval2015-tju} &  0.285     & 0.8           &  0.49        \\
    \midrule
    NII-UIT \cite{mediaeval2015-nii} &  0.268     & 0.6           &  0.44        \\
    \midrule
    \emph{this work}    &  \textbf{0.303}     & \textbf{1.0}           &  \textbf{0.55}  \\            
    \bottomrule
    \end{tabular}%
  }
\end{table}%